\documentclass[prd,aps,showpacs,preprintnumbers,amssymb]{revtex4}

\usepackage{color}
\usepackage{epsf}
\usepackage{slashed}

\begin{document}
\title{%
\hfill{\normalsize\vbox{%
\hbox{}
 }}\\
{ Generalized fermion symmetry, its currents algebra and   Ward-Takahashi identities}}

\author{Amir H. Fariborz
$^{\it \bf a}$~\footnote[1]{Email:
 fariboa@sunyit.edu}}

\author{Renata Jora
$^{\it \bf b}$~\footnote[2]{Email:
 rjora@theory.nipne.ro}}

\affiliation{$^{\bf \it a}$ Department of Matemathics/Physics, SUNY Polytechnic Institute, Utica, NY 13502, USA}
\affiliation{$^{\bf \it b}$ National Institute of Physics and Nuclear Engineering PO Box MG-6, Bucharest-Magurele, Romania}

\date{\today}

\begin{abstract}

We introduce a new local symmetry into the fermion sector of a gauge invariant Lagrangian which may or may not contain a scalar or spontaneous symmetry breaking.  The Standard Model in the  unitary gauge and QCD  are particular cases where this symmetry may apply. We determine the associated vector and axial vector currents and their conservation laws. We show that a single current conservation law may lead to multiple Ward-Takahashi identities.  Our results can potentially have  important consequences for
effective models of low-energy QCD and hadron structure.   As an specific example, we discuss the construction of tetraquark states within a generalized linear sigma model and show that this new symmetry probes the tetraquarks in a manner that is consistent with the large $N_c$ limit of QCD.

\end{abstract}
\maketitle

\section{Introduction}

Gauge  invariant theories with fermions and Yukawa couplings stand at the base of the standard model of elementary particles and QCD. The gauge symmetry is implemented for all particles in the Lagrangian according to their group representations. There may also be global symmetries associated to each particular Lagrangian. Then one can derive the conserved currents and charges and the corresponding Ward-Takahashi  identities \cite{Ward}, \cite{Takahashi}. It was shown in  \cite{Jora} that it is always possible to find out new symmetries of a gauge invariant theory  that may represent combinations of the old ones. Here we shall expand  and improve the point of view introduced in \cite{Jora} to present and discuss  a new symmetry applicable to the fermion sector of any Lagrangian and in particular to that of the electroweak theory or QCD. This symmetry extends our present knowledge of the partition function through new vector and axial vector currents conserved or anomalous and may have important consequences. Specifically we discuss the case of a low energy QCD Lagrangian, a generalized linear sigma model with two chiral meson nonets, one with a quark antiquark structure the other one with a four quark composition.  The symmetry at hand distinguishes among three possible tetraquark structures that otherwise behave completely identical under the chiral $U(3)_L \times U(3)_R$. Moreover the new symmetry may be used to construct a more comprehensive effective QCD Lagrangian that contains, besides scalars and pseudoscalars also vectors, pseudovectors and tensors and may give a hint with regard to the actual hadron composition.

Section II introduces the new symmetry for  a gauge abelian model with fermions coupled to scalars through Yukawa terms.  In section II we apply the Fujikawa \cite{Fujikawa} method to determine the behavior of currents. Sector II contains a generalization of the symmetry for the more intricate fermion sector of the standard model. In section IV we determine simple Ward-Takahashi identities associated to the new symmetry. Section V is dedicated to Conclusions.

\section{A new symmetry that reinforces an old one}
We consider a $U(1)$ gauge model with fermions and scalars where the scalar might be charged  under the gauge group but couples to the fermions through Yukawa interactions. The Lagrangian of interest has the expression,
\begin{eqnarray}
{\cal L}=i\bar{\Psi}\gamma^{\mu}D_{\mu}\Psi+yB\bar{\Psi}\Psi.
\label{lagr546738}
\end{eqnarray}
where we took into account only the fermion sector and worked in the unitary gauge where the Goldstone boson was eliminated for the case where the scalar field was charged under the gauge group.
Here $B$ is the scalar field, $y$ is the Yukawa coupling and,
\begin{eqnarray}
D_{\mu}\Psi=(\partial_{\mu}-igA_{\mu}\partial_{\mu})\Psi.
\label{covder546378}
\end{eqnarray}
First we will show that the above portion of the Lagrangian and consequently the full Lagrangian of the theory is invariant under the transformation (see \cite{Jora} where we introduced a similar symmetry):
\begin{eqnarray}
&&\Psi \rightarrow \Psi + k(\gamma^{\rho}D_{\rho}\Psi-iyB\Psi)
\nonumber\\
&&\bar{\Psi}\rightarrow \bar{\Psi}+k\partial_{\rho}\bar{\Psi}\gamma^{\rho}+ikg\bar{\Psi}\gamma^{\rho}\Psi A_{\rho}+iyB\bar{\Psi},
\label{res638274651}
\end{eqnarray}
where $k$ is a parameter with mass dimension $m^{-1}$.
We start with,
\begin{eqnarray}
\delta {\cal L}&=&i\delta(\bar{\Psi})\gamma^{\mu}D_{\mu}\Psi+i\bar{\Psi}\gamma^{\mu}D_{\mu}\delta(\Psi)+yB\delta (\bar{\Psi})\Psi+yB\bar{\Psi}\delta{\Psi}=
\nonumber\\
&&ik(\partial_{\rho}\bar{\Psi})\gamma^{\rho}\gamma^{\mu}D_{\mu}\Psi-kg\bar{\Psi}\gamma^{\rho}\gamma^{\mu}D_{\mu}\Psi A_{\rho}-ykB\bar{\Psi}\gamma^{\mu}D_{\mu}\Psi+
\nonumber\\
&&ik\bar{\Psi}\gamma^{\mu}D_{\mu}(\gamma^{\rho}D_{\rho}-iyB)\Psi+kyB\partial_{\mu}\bar{\Psi}\gamma^{\mu}\Psi+ikg\bar{\Psi}\gamma^{\mu}\Psi A_{\mu}B+iy^2B^2\bar{\Psi}\Psi+
\nonumber\\
&&kyB\bar{\Psi}\gamma^{\mu}\partial_{\mu}\Psi-ikgB\bar{\Psi}\gamma^{\mu}\Psi A_{\mu}-iy^2B^2\bar{\Psi}\Psi=
\nonumber\\
&&ik\partial_{\rho}(\bar{\Psi}\gamma^{\rho}\gamma^{\mu}D_{\mu}\Psi)-
ik\bar{\Psi}\gamma^{\rho}\gamma^{\mu}(\partial_{\rho}-igA_{\rho})D_{\mu}\Psi+ik\bar{\Psi}\gamma^{\mu}D_{\mu}\gamma^{\rho}D_{\rho}\Psi-
\nonumber\\
&&kyB\bar{\Psi}\gamma^{\mu}D_{\mu}\Psi+kyB\bar{\Psi}\gamma^{\mu}D_{\mu}\Psi+
\nonumber\\
&&ky\bar{\Psi}\gamma^{\mu}\partial_{\mu}B\Psi+kyB\partial_{\mu}\gamma^{\mu}\Psi+kyB(\partial_{\mu}\bar{\Psi})\gamma^{\mu}\Psi=
\nonumber\\
&&ik\partial_{\rho}(\bar{\Psi}\gamma^{\rho}\gamma^{\mu}D_{\mu}\Psi)+ky\partial_{\rho}(\bar{\Psi}\gamma^{\rho}\Psi B).
\label{ofetxre637489567}
\end{eqnarray}
Then the conserved current is:
\begin{eqnarray}
J_{\rho}=i(\bar{\Psi}\gamma^{\rho}\gamma^{\mu}D_{\mu}\Psi)+y(\bar{\Psi}\gamma^{\rho}\Psi B).
\label{res7253647}
\end{eqnarray}
In a similar way it can be shown (just note that the terms that contain anticommutators contain an $i$)  that also the transformation,
\begin{eqnarray}
&&\Psi \rightarrow \Psi + ik(\gamma^5\gamma^{\rho}D_{\rho}\Psi-iy\gamma^5B\Psi)
\nonumber\\
&&\bar{\Psi}\rightarrow \bar{\Psi}+ik\partial_{\rho}\bar{\Psi}\gamma^{\rho}\gamma^5-kg\bar{\Psi}\gamma^{\rho}\gamma^5 A_{\rho}-kyB\bar{\Psi}\gamma^5,
\label{res63827465}
\end{eqnarray}
is also a symmetry of the Lagrangian with the variation given by:
\begin{eqnarray}
\delta{\cal L}=-k\partial_{\rho}(\bar{\Psi}\gamma^{\rho}\gamma^5\gamma^{\mu}D_{\mu}\Psi)+iky\partial_{\rho}(B\bar{\Psi}\gamma^{\rho}\gamma^5\Psi).
\label{trlagr463784956768}
\end{eqnarray}
The associated axial current is:
\begin{eqnarray}
K_{\rho}=-(\bar{\Psi}\gamma^{\rho}\gamma^5\gamma^{\mu}D_{\mu}\Psi)+iy(B\bar{\Psi}\gamma^{\rho}\gamma^5\Psi).
\label{axialcurrent53856478}
\end{eqnarray}
 We can  check whether these symmetries are anomalous by applying the Fujikawa method \cite{Fujikawa}.

\section{Applying Fujikawa method}

Here the transformation in Eq. (\ref{res638274651}) is regarded as a change of variable in the partition function for the fermion Lagrangian. We expand the fermions wave functions in eigenstates of the Hamiltonian  $\Phi_n$ and $\Phi_n^{\dagger}$ with the property $(\gamma^{\mu}D_{\mu}-iyB)\Phi_n=\lambda_n\Phi_n$:
\begin{eqnarray}
&&\Psi=\sum_n b_n \Phi_n
\nonumber\\
&&\bar{\Psi}=\sum_n \bar{b}_n\Phi_n^{\dagger}
\nonumber\\
&&\Psi'=\sum b_n'\Phi_n
\nonumber\\
&&\bar{\Psi}'=\sum\bar{b}_n'\Phi_n^{\dagger},
\label{res745364789}
\end{eqnarray}
where $\Psi'$ and $\bar{\Psi}'$ are the fields transformed under Eq. (\ref{res638274651}).
This yields in the standard Fujikawa approach:
\begin{eqnarray}
&&b_m'=\sum_nb_n\int d^4 x\Phi_m^{\dagger}(x)[1+k\gamma^{\mu}\partial_{\mu}-ig\gamma^{\mu}A_{\mu}-iyB]\Phi_n(x)
\nonumber\\
&&\bar{b}_m'=\sum \bar{b}_n\int d^4 x \Phi_n^{\dagger}(x)[1-k\gamma^{\mu}\partial_{\mu}+igA_{\mu}\gamma^{\mu}+iyB]\Phi_m(x).
\label{res745385647}
\end{eqnarray}
Here we integrated by parts in the first term of the second line.

Since $d\bar{\Psi}d\Psi=\prod_n d\bar{b_n}b_n$ and $d\bar{\Psi}'d\Psi'=\prod_n d \bar{b}_n'db_n'$ the transformation jacobian is written in terms of  the product of determinants:
\begin{eqnarray}
{\cal J}^{-1}=\det[C^m_n]\det[C^{m\prime}_n]
\label{jacobr65748}
\end{eqnarray}
where,
\begin{eqnarray}
&&C^m_n=[\delta^m_n+k\int d^4 x\Phi_m^{\dagger}[\gamma^{\mu}\partial_{\mu}-ig\gamma^{\mu}A_{\mu}-iyB]\Phi_n(x)]
\nonumber\\
&&C_n^{\prime m}=[\delta^m_n-k\int d^4 x\Phi_n^{\dagger}[\gamma^{\mu}\partial_{\mu}-ig\gamma^{\mu}A_{\mu}-iyB]\Phi_m(x)]
\label{res72538956}
\end{eqnarray}
Then it is obvious that the two determinants cancel each other in first order of $k$ so they do not bring any contributions.

Similarly one can show that the jacobian associated with the transformation in Eq. (\ref{res63827465}) is:
\begin{eqnarray}
&&{\cal J}^{-1}=\det[D^m_n]\det[D^{m\prime}_n]=
\nonumber\\
&&\det[\delta^m_n+\int d^4x k\Phi_m^{\dagger}[i\gamma^5\gamma^{\mu}\partial_{\mu}+g\gamma^5\gamma^{\mu}A_{\mu}+y\gamma^5B]\Phi_m]\times
\nonumber\\
&&\det[\delta^m_n+\int d^4 xk \Phi_n^{\dagger}[-i\gamma^{\mu}\gamma^5\partial_{\mu}-g\gamma^{\mu}\gamma^5A_{\mu}-y\gamma^5B]\Phi_m].
\label{ax6475849}
\end{eqnarray}
One can further write in first order:
\begin{eqnarray}
\det[D^m_n]\det[D^{m\prime}_n]=\exp\Bigg[{\rm Tr}2\int d^4x k\Phi_m^{\dagger}[i\gamma^5\gamma^{\mu}\partial_{\mu}+g\gamma^5\gamma^{\mu}A_{\mu}]\Phi_m\Bigg]
\label{res75846599}
\end{eqnarray}
In order to compute the above expression we need to regularize it. For that we sandwich between the eigenstates the operator $\exp[\frac{\lambda_n^2}{M^2}]$ where $\lambda_n$ are the eigenvalues and in the end one takes the limit $M\rightarrow \infty$. Since by the orthonormalization of the states $\Phi_n$ we will need to take a trace of the operator between the eigenstates we observe that we get a contribution different than zero only if we expand the exponential in the second order and even then only some of the terms contribute. Finally we need to determine:
\begin{eqnarray}
{\rm Tr} \Bigg[\frac{2}{i}\int d^4 x \Phi_m^{\dagger}\gamma^5(i\slashed{D})\exp[\frac{(i\slashed{D}+yB)^2}{M^2}]\Phi_m(x)\Bigg]
\label{ax746358}
\end{eqnarray}
which leads to:
\begin{eqnarray}
&&{\rm Tr}\Bigg[\frac{1}{i}\gamma^5(i\slashed{D})y[(i\slashed{D})^3(i\slashed{D}B)+i\slashed{D}B(i\slashed{D})^2+(i\slashed{D})^2B(i\slashed{D})+B(i\slashed{D})^3]\Bigg]=
\nonumber\\
&&=-4\epsilon^{\mu\nu\rho\sigma}[D_{\mu}D_{\nu}D_{\rho}D_{\sigma}B+D_{\mu}D_{\nu}BD_{\rho}D_{\sigma}+D_{\mu}D_{\nu}D_{\rho}BD_{\sigma}+D_{\mu}BD_{\nu}D_{\rho}D_{\sigma}]=
\nonumber\\
&&-4\epsilon^{\mu\nu\rho\sigma}\Bigg[\frac{1}{2}(-ig)F_{\mu\nu}\frac{1}{2}(-ig)F_{\rho\sigma}B+\frac{1}{2}(-ig)F_{\mu\nu}\frac{1}{2}(-ig)F_{\rho\sigma}B+
\nonumber\\
&&\frac{1}{2}(-ig)F_{\mu\nu}D_{\rho}BD_{\sigma}+\frac{1}{2}(-ig)D_{\mu}BD_{\sigma}F_{\rho\sigma}\Bigg].
\label{res624356}
\end{eqnarray}
We shall analyze in more detail the last two terms in the Eq. (\ref{res624356}):
\begin{eqnarray}
&&-4\epsilon^{\mu\nu\rho\sigma}[\frac{1}{2}(-ig)F_{\mu\nu}D_{\rho}BD_{\sigma}+\frac{1}{2}(-ig)D_{\mu}BD_{\sigma}F_{\rho\sigma}]
\nonumber\\
&&+2ig\epsilon^{\mu\nu\rho\sigma}[F_{\mu\nu}(\partial_{\rho}-igA_{\rho})B(\partial_{\sigma}-igA_{\sigma})+
(\partial_{\mu}-igA_{\mu})B(\partial_{\nu}-igA_{\nu})F_{\rho\sigma}]=
\nonumber\\
&&+2ig\epsilon^{\mu\nu\rho\sigma}[F_{\mu\nu}(-igA_{\sigma}\partial_{\rho}B-igB\partial_{\rho}A_{\sigma})+(-igA_{\nu}\partial_{\mu}B-igB\partial_{\mu}A_{\nu})F_{\rho\sigma}]=
\nonumber\\
&&2g^2\epsilon^{\mu\nu\rho\sigma}[BF_{\mu\nu}F_{\rho\sigma}+2A_{\sigma}\partial_{\rho}BF_{\mu\nu}]
\label{twofinal547839}
\end{eqnarray}
The final result computed from Eqs. (\ref{res624356}) and (\ref{twofinal547839}) is:
\begin{eqnarray}
&&{\rm Tr}\Bigg[\frac{1}{i}\gamma^5(i\slashed{D})y[(i\slashed{D})^3(i\slashed{D}B)+(i\slashed{D}B)(i\slashed{D})^2+(i\slashed{D})^2B(i\slashed{D})+B(i\slashed{D})^3]\Bigg]=
\nonumber\\
&&4g^2\epsilon^{\mu\nu\rho\sigma}[F_{\mu\nu}F_{\rho\sigma}B+F_{\mu\nu}A_{\sigma}\partial_{\rho}B]
\label{res738456378}
\end{eqnarray}
Since the partial derivatives should be consider separately in the exponential (the above results exclude the simple derivatives) the result of integration is (see \cite{Peskin}):
\begin{eqnarray}
\langle 0| \exp[\frac{-\partial^2}{M^2}|0\rangle=i\frac{M^4}{16\pi^2}
\label{res6384657}
\end{eqnarray}
and moreover the jacobian is at the power $-1$ the final contribution of the axial transformation is:
\begin{eqnarray}
&&\exp\Bigg[-i\int d^4 xk(x) [\partial_{\rho}(\bar{\Psi}\gamma^{\rho}\gamma^5\gamma^{\mu}D_{\mu}\Psi -iy\partial_{\rho}(B\bar{\Psi}\gamma^{\rho}\gamma^5\Psi)]-
\nonumber\\
&&4ig^2\frac{1}{16\pi^2}\epsilon^{\mu\nu\rho\sigma}\int d^4 x k(x)[F_{\mu\nu}F_{\rho\sigma}B+F_{\mu\nu}A_{\sigma}\partial_{\rho}B]\Bigg].
\label{res6496758}
\end{eqnarray}
This yields the anomalous conservation of the axial current (see Eq. (\ref{axialcurrent53856478})):
\begin{eqnarray}
&&\partial_{\rho}K^{\rho}=[\partial_{\rho}(\bar{\Psi}\gamma^{\rho}\gamma^5\gamma^{\mu}D_{\mu}\Psi -iy\partial_{\rho}(B\bar{\Psi}\gamma^{\rho}\gamma^5\Psi)]=
\nonumber\\
&&-\frac{4g^2}{16\pi^2}\epsilon^{\mu\nu\rho\sigma}[F_{\mu\nu}F_{\rho\sigma}B+F_{\mu\nu}A_{\sigma}\partial_{\rho}B].
\label{finalres73546789}
\end{eqnarray}

 Note that the expression in Eq. (\ref{finalres73546789}) can be made gauge invariant up to a total derivative.

It is important to relate the axial anomaly that we obtained in Eq. (\ref{finalres73546789}) with the regular axial anomaly given by the divergence of the current $J^a_{\rho}=\bar{\Psi}\gamma^{\rho}\gamma^{5}\Psi$.  This anomaly is well understood in two, three or four dimensions and stems from the regularization of UV infinite integrals in the Feynman diagram or Fujikawa approaches.  The axial anomaly depicted in Eq. (\ref{finalres73546789}) is strongly related to the standard axial anomaly and has exactly the same origin; the regularization of UV divergences or more directly  fermion triangle diagrams with the scalar attached at one vertex. This can be also seen schematically from the current divergence that we calculate. Then  the axial anomaly introduced in Eq. (\ref{finalres73546789}) is a straightforward generalization  of the regular axial anomaly and may serve not only to computing quantities in perturbation theory but also for computing higher order anomalous contributions in low-energy QCD effective theories like the linear sigma model or chiral perturbation theory through the introduction of new coupling of axial type.  A more detail discussion of this aspect will be made in the last section.

\section{The new symmetries applied to the standard model}

Instead of proving that the symmetry introduced in section II works also for the nonabelian case we will show that it is true for the more intricate case of the standard model. For that we pick arbitrarily one quark doublet
and the corresponding part of the Lagrangian again in the unitary gauge where the Goldstone bosons are eliminated from the theory \cite{Cheng}:
\begin{eqnarray}
{\cal L}_2&=&
\bar{q}_Li(\slashed{\partial}-i\frac{g}{2}\tau \slashed{A}-i\frac{g'}{6}\slashed{B})q_L+\bar{p}_Ri(\slashed{\partial}-2i\frac{g'}{3}\slashed{B})p_R+
\nonumber\\
&&\bar{n}_Ri(\slashed{\partial}+i\frac{g'}{3}\slashed{B})n_R+y_pH(\bar{p}_Rp_L+\bar{p}_Lp_R)+y_nH(\bar{n}_Rn_l+\bar{n}_Ln_R).
\label{res72536789}
\end{eqnarray}
Here $A^i_{\mu}$ are the $SU(2)_L$ gauge fields, $B_{\mu}$ is the $U(1)_Y$ gauge field and $H$ is the Higgs boson.
The transformation of interest is:
\begin{eqnarray}
&&p_L\rightarrow p_L+k(\slashed{\partial}-2i\frac{g}{3}\slashed{B})p_r-iy_pHp_L
\nonumber\\
&&\bar{p}_L\rightarrow\bar{p}_L+k\bar{p}_R(\overleftarrow{\slashed{\partial}}+2i\frac{g'}{3}\slashed{B})+iy_pH\bar{p}_L
\nonumber\\
&&p_R\rightarrow p_R +k(\slashed{\partial}-i\frac{g}{2}\tau\slashed{A}-i\frac{g'}{6}\slashed{B})_{1j}q_{lj}
\nonumber\\
&&\bar{p}_R\rightarrow \bar{p}_R+k\bar{q}_{Lj}(\overleftarrow{\slashed{\partial}}+i\frac{g}{2}\tau \slashed{A}+i\frac{g'}{6}\slashed{B})_{j1}+iy_pH\bar{p}_R.
\label{transuty76}
\end{eqnarray}

Then the variation of the Lagrangian in Eq. (\ref{res72536789}) under the transformation in Eq. (\ref{transuty76}) is given by:
\begin{eqnarray}
\delta{\cal L}_2&=&
\delta(\bar{p}_L)i(\slashed{\partial}-i\frac{g}{2}\tau \slashed{A}-i\frac{g'}{6}\slashed{B})_{1j}q_{Lj}+
\bar{q}_{Lj}i(\slashed{\partial}-i\frac{g}{2}\tau \slashed{A}-\frac{g'}{6}\slashed{B})_{j1}(\delta p_L)+
\nonumber\\
&&\delta(\bar{p}_R)i(\slashed{\partial}-2i\frac{g'}{3}\slashed{B})p_R+
\bar{p}_Ri(\slashed{\partial}-2i\frac{g'}{3}\slashed{B})\delta p_R+
\nonumber\\
&&y_pH[\delta(\bar{p}_R)p_L+\bar{p}_R\delta p_L +\delta(\bar{p}_L)p_R+\bar{p}_L\delta p_R].
\label{firs726588}
\end{eqnarray}
We shall consider first in all terms the part of the transformation that contains the scalar $H$, ie.:
\begin{eqnarray}
&&k\Bigg[H\bar{q}_{Li}(\overleftarrow{\slashed{\partial}}+\frac{g}{2}\tau \slashed{A}+i\frac{g'}{6}\slashed{B})_{i1}p_L+\bar{q}_{Li}(\slashed{\partial}-i\frac{g}{2}\tau\slashed{A}-i\frac{g'}{6}\slashed{B})_{i1}(Hp_L)+
\nonumber\\
&&H\bar{p}_L(\slashed{\partial}-i\frac{g}{2}\tau\slashed{A}-i\frac{g'}{6}\slashed{B})_{1j}q_{Lj}-H\bar{p}_L(\slashed{\partial}-i\frac{g}{2}\tau\slashed{A}-i\frac{g'}{6}\slashed{B})_{1j}q_{Lj}+
\nonumber\\
&&H\bar{p}_R(\overleftarrow{\slashed{\partial}}+2i\frac{g'}{3}\slashed{B})p_R+\bar{p}_R(\slashed{\partial}-2i\frac{g'}{3}\slashed{B})(Hp_R)+
\nonumber\\
&&H\bar{p}_R(\slashed{\partial}-2i\frac{g'}{3}\slashed{B})p_R-H\bar{p}_R(\slashed{\partial}-2i\frac{g'}{3}\slashed{B})p_R\Bigg]=
\nonumber\\
&&k\partial_{\mu}(\bar{p}_L\gamma^{\mu}p_L H+\bar{p}_R\gamma^{\mu}p_RH).
\label{term453748}
\end{eqnarray}
Here the left arrow derivative refers only to the fermions and all other simpler contributions that contain $H$ and cancelled are not mentioned.

Next step is to consider the terms in the variation that do not contain $H$:
\begin{eqnarray}
&&i\bar{p}_R(\overleftarrow{\slashed{\partial}}+2i\frac{g'}{3}\slashed{B})(\slashed{\partial}-i\frac{g}{2}\slashed{A}-i\frac{g'}{6}\slashed{B})_{1j}q_{Lj}+
\nonumber\\
&&i\bar{p}_R(\slashed{\partial}-2i\frac{g'}{3}\slashed{B})(\slashed{\partial}-i\frac{g}{2}\tau\slashed{A}-i\frac{g'}{6}\slashed{B})_{1j}q_{Lj}+
\nonumber\\
&&i\bar{q}_{Lj}(\slashed{\partial}-i\frac{g}{2}\tau\slashed{A}-i\frac{g'}{6}\slashed{B})_{j1})(\slashed{\partial}-2i\frac{g'}{3}\slashed{B})p_R+
\nonumber\\
&&i\bar{q}_{Lj}(\overleftarrow{\slashed{\partial}}+i\frac{g}{2}\tau\slashed{A}+i\frac{g'}{6}\slashed{B})_{j1}(\slashed{\partial}-2i\frac{g'}{3}\slashed{B})p_R=
\nonumber\\
&&i\partial_{\mu}[\bar{p}_R\gamma^{\mu}(\slashed{\partial}-i\frac{g}{2}\tau\slashed{A}-i\frac{g'}{6}\slashed{B})_{1k}q_{Lk}+\bar{p}_L\gamma^{\mu}(\slashed{\partial}-2i\frac{g'}{3}\slashed{B})p_R].
\label{rezyruti75}
\end{eqnarray}
The divergence of currents is obtained by adding Eqs. (\ref{term453748}) and (\ref{rezyruti75}):
\begin{eqnarray}
\partial_{\mu}J^{\mu}&=&\partial_{\mu}(\bar{p}_L\gamma^{\mu}p_L H+\bar{p}_R\gamma^{\mu}p_RH)+
\nonumber\\
&&i\partial_{\mu}[\bar{p}_R\gamma^{\mu}(\slashed{\partial}-i\frac{g}{2}\tau\slashed{A}-i\frac{g'}{6}\slashed{B})_{1k}q_{Lk}+\bar{p}_L\gamma^{\mu}(\slashed{\partial}-2i\frac{g'}{3}\slashed{B})p_R],
\label{cons647382}
\end{eqnarray}
which leads to the current:
\begin{eqnarray}
J^{\mu}=(\bar{p}_L\gamma^{\mu}p_L H+\bar{p}_R\gamma^{\mu}p_RH)+i[\bar{p}_R\gamma^{\mu}(\slashed{\partial}-i\frac{g}{2}\tau\slashed{A}-i\frac{g'}{6}\slashed{B})_{1k}q_{Lk}+\bar{p}_L\gamma^{\mu}(\slashed{\partial}-2i\frac{g'}{3}\slashed{B})p_R].
\label{currents546}
\end{eqnarray}
However we expect that this current is anomalous since it contains axial terms. Note that similar currents and conservation laws can be obtained for all standard model fermions. Whether this anomaly has any physical significance is a non-trivial question to answer and requires further analysis.

\section{Ward-Takahashi identities}

Here we will obtain simple Ward Takhashi identities for the abelian Higgs model with fermions in the standard approach. Thus we consider the invariance of the Lagrangian under the symmetry stated in Eq. (\ref{res63827465}) and with the variation of the Lagrangian given in Eq.(\ref{ofetxre637489567}). Then the following identity holds:
\begin{eqnarray}
&&\frac{1}{Z}\int d\bar{\Psi} d \Psi \exp[i\int d^4 x {\cal L}]\Psi(x_1)\bar{\Psi}(x_2)=
\nonumber\\
&&\frac{1}{Z}\int d \bar{\Psi}' d\Psi'  \exp[i\int d^4 x {\cal L}']\Psi'(x_1)\bar{\Psi}'(x_2),
\label{id6475867}
\end{eqnarray}
which leads to,
\begin{eqnarray}
0&=&\frac{1}{Z}\int d \bar{\Psi} d\Psi \Bigg[[\int d^4 x [-k(x)\partial_{\rho}(\bar{\Psi}\gamma^{\rho}\gamma^{\mu}D_{\mu}\Psi(x)+ik(x)y\partial_{\rho}(\bar{\Psi}\gamma^{\rho}\Psi B)]\Psi(x_1)\bar{\Psi}(x_2)+
\nonumber\\
&&\int d^4 x k(x_1)\delta(x-x_1)[\gamma^{\mu}\partial_{\mu}\Psi(x_1)-ig\gamma^{\mu}\Psi(x_1)A_{\mu}(x_1)-iyB(x_1)\Psi(x_1)]\bar{\Psi}(x_2)+
\nonumber\\
&&\int d^4 x k(x_2)\delta(x-x_2)\Psi(x_1)[\partial_{\mu}\bar{\Psi}(x_2)\gamma^{\mu}+ig\bar{\Psi}(x_2)\gamma^{\mu}A_{\mu}(x_2)+iyB(x_2)\bar{\Psi}(x_2)\Bigg] \exp[i\int d^4 x {\cal L}].
\label{res634578}
\end{eqnarray}
Here $Z$ is the partition function.
We can further process Eq. (\ref{res634578}) to obtain:
\begin{eqnarray}
&&\langle 0 |T([-k(x)\partial_{\rho}(\bar{\Psi}\gamma^{\rho}\gamma^{\mu}D_{\mu}+ik(x)y\partial_{\rho}(\bar{\Psi}\gamma^{\rho}\Psi B)]\Psi(x_1)\bar{\Psi}(x_2)|0\rangle=
\nonumber\\
&&-\langle 0|\Bigg[\int d^4 x k(x_1)\delta(x-x_1)[\gamma^{\mu}\partial_{\mu}\Psi(x_1)-ig\gamma^{\mu}\Psi(x_1)A_{\mu}(x_1)-iyb(x_1)\Psi(x_1)]\bar{\Psi}(x_2)+
\nonumber\\
&&\int d^4 x k(x_2)\delta(x-x_2)\Psi(x_1)[\partial_{\mu}\bar{\Psi}(x_2)\gamma^{\mu}+ig\bar{\Psi}(x_2)\gamma^{\mu}A_{\mu}(x_2)+iyB(x_2)\bar{\Psi}(x_2)\Bigg]|0\rangle.
\label{res637489}
\end{eqnarray}
There are multiple Ward-Takahashi identities that one can obtain from Eq. (\ref{res637489}). We shall consider one simple case and briefly discuss another. First we consider the Fourier transforms of the relation in  Eq. (\ref{res637489}) then introduce a factor of $\delta(x)$ and integrate over $x$. Furthermore the field $B$ is replaced by its vacuum expectation value and the terms that contain $A_{\mu}$ are set to zero. We denote the renormalization constant for the wave function of fermion by $Z_2$, that of the fermion gauge field vertex $Z_1$, that of the fermion scalar field vertex $Z_y$ and finally  that of the vacuum expectation value $Z_v$. We obtain,
\begin{eqnarray}
&&i(p_1-p_2)_{\rho}ip_{2\mu}S(p_1)\gamma^{\rho}\gamma^{\mu}S(p_2)\Gamma_2+(p_1-p_2)_{\rho}yS(p_1)\gamma^{\rho}S(p_2)\Gamma_m=
\nonumber\\
&&-ip_{2\mu}\gamma^{\mu}S(p_2)+ip_{1\mu}S(p_1)\gamma^{\mu}+iy_0v_0\gamma^{\mu}S(p_2)-iy_0v_0S(p_1)\gamma^{\mu}.
\label{firstres6478859}
\end{eqnarray}
Here $p_1$ and $p_2$ come from the Fourier transform of $\Psi(x_1)$ and $\bar{\Psi}(x_2)$ and $S(r)$ is the all order propagator of the fermion field. By considering $p_1=k+p_2$ and  $p_2=p$
we divide  to the left and right by the propagators $S(p_1)$ and $S(p_2)$ which yields:
\begin{eqnarray}
-k_{\rho}\gamma^{\rho}(p_{\mu}\gamma^{\mu}\Gamma_2-Z_v^{-1}\Gamma_m)=-iS(p+k)^{-1}[\gamma^{\mu}p_{\mu}-y_0v_0]+i((p+k)_{\mu}\gamma^{\mu}-y_0v_0)S(p)^{-1}.
\label{res73647}
\end{eqnarray}
Next we set $p$ near mass shell $S(p)\approx \frac{Z_2}{p-m}$ and expand in $k$:
\begin{eqnarray}
-k^{\rho}\gamma^{\rho}(p_{\mu}\gamma^{\mu}\frac{1}{Z_2}-y_0v_0\frac{1}{Z_mZ_v})=-k^{\rho}\gamma^{\rho}\frac{1}{Z_2}(p_{\mu}\gamma^{\mu}-y_0v_0).
\label{res63788}
\end{eqnarray}
Here we used the expression of the renormalized mass term as being $m=\frac{m_0}{Z_m}$, that of the kinetic term as being $\frac{p_{\mu}}{Z_2}$  and $v=Z_v^{-1}v_0$. Then we obtain the identity,
\begin{eqnarray}
Z_m=Z_v^{-1}Z_2
\label{res647895}
\end{eqnarray}
which is already settled in the literature \cite{Machacek1}-\cite{Bohm}. By simply introducing in Eq. (\ref{id6475867}) a gauge field $A_{\sigma}(x_3)$ one can obtain in the same simple manner the standard Ward Takahashi equality $Z_2=Z_1$. Note that all relation between the renormalization constants can be retrieved from  a single  law of conservation in the partition function.

\section{Discussion and conclusions}

In this work we introduced and discussed a new symmetry applied to the fermions in a gauge invariant Lagrangian. This symmetry is  related to the global symmetries and symmetry under translations with two main differences: the Lagrangian is symmetric   under the transformation of each flavor of fermions separately  and this applies also to more intricate structures like that of the standard model: the symmetry includes in its expression both the gauge and the scalar fields that are coupled to the fermions.  We further determined the vector current and showed that it is conserved in each order of perturbation theory and the axial vector currents for which we calculated the inherent anomaly. Because of the complexity of the symmetry considered it was very amenable to obtain Ward-Takahashi identities associated to the gauge fermion and scalar fermion vertices.

In general the global and local symmetry associated to each Lagrangian are known from its construction. Besides the well known symmetries is it always possible to find combinations of them that may lead to new currents algebra  at the level of the Lagrangian and to more intricate Ward-Takahashi identities at the quantum level. In this work we introduced such a symmetry for the fermions sector of any Lagrangian. The results obtained here
may be useful in calculating processes and correlators  both in QCD or the standard model and to determine new possible conserved quantities that may have phenomenological implications.

A potential application can be in study of nontrivial meson fields (such as tetraquarks) and their construction in terms of the  underlying quark fields.  As an example, consider the generalized linear sigma model with two chiral nonets one with a quark-antiquark structure the other one with a four-quark content. The model was introduced in \cite{Jora1} and further discussed in \cite{Jora2}-\cite{Jora4}.
The quark-antiquark chiral nonet has a simple structure in terms of quark fields:
\begin{eqnarray}
M^b_a=(q_{bA})^{\dagger}\gamma_4\frac{1+\gamma_5}{2}q_{aA}= S_a^b +i\phi_a^b,
\label{non546788}
\end{eqnarray}
where the small (capital) letters are the flavor (color) indices and $S$ ($\phi$) are the quark-antiquark scalar (pseudoscalar) meson nonets.

Unlike the simple quark substructure in (\ref{non546788}),   to write the quark schematic composition for the tetraquark fields, there are three possibilities that are  compatible with the global symmetries:

\begin{eqnarray}
&&(a) \hskip .5cm M_a^{(2)b}=\epsilon_{acd}\epsilon^{bef}(M^{\dagger})^c_e(M^{\dagger})^d_f
\nonumber\\
&&(b) \hskip .5cm M_g^{(3)f}=(L^{gA})^{\dagger}R^{fA}
\nonumber\\
&&(c) \hskip .5cm M_g^{(4)f}=(L^{g}_{\mu\nu,AB})^{\dagger}R^{f}_{\mu\nu,AB}
\label{tetraq_abc}
\end{eqnarray}
where
\begin{eqnarray}
&&L^{gE}=\epsilon^{gab}\epsilon^{EAB}q^T_{aA}C^{-1}\frac{1+\gamma_5}{2}q_{bB}
\nonumber\\
&&R^{gE}=\epsilon^{gab}\epsilon^{EAB}q^T_{aA}C^{-1}\frac{1-\gamma_5}{2}q_{bB}
\nonumber\\
&&L^{g}_{\mu\nu,AB}=\epsilon^{gab}q^T_{aA}C^{-1}\sigma_{\mu\nu}\frac{1+\gamma_5}{2}q_{bB}
\nonumber\\
&&R^{g}_{\mu\nu,AB}=\epsilon^{gab}q^T_{aA}C^{-1}\sigma_{\mu\nu}\frac{1-\gamma_5}{2}q_{bB},
\label{res7466578}
\end{eqnarray}
where $C$ is the charge conjugation.
The first  composite field ($a$) in (\ref{tetraq_abc}) is of the ``molecular'' type (two color singlet quark-antiquark meson $M$ bound together to form a tetraquark) whereas the second and third substructures ($b$ and $c$) are of the form diquark-antidiquark compositions (where in $b$ and $c$ the diquarks are in spin 0 and 1, respectively).     Fierz transfromations can establish a linear relationship between the three combination in (\ref{tetraq_abc}).   As a result,   the tetraquark composite field $M'$ is, in general,  a linear combination of any two of the three substructures in (\ref{tetraq_abc}).   In addition,
\begin{equation}
M^\prime = S^\prime +i\phi^\prime,
\label{sandphi1}
\end{equation}
where $S'$ ($\phi'$) is the tetraquark scalar (pseudoscalar) meson nonet.
The $3 \times 3$ chiral nonets $M$ and $M'$ transform in the same way under the SU(3) chiral symmetry,
\begin{eqnarray}
M &\rightarrow& U_L\, M \, U_R^\dagger,\nonumber\\
M' &\rightarrow& U_L\, M' \, U_R^\dagger,
\end{eqnarray}
but transform differently under U(1)$_{\rm A}$
transformation properties
\begin{eqnarray}
M &\rightarrow& e^{2i\nu}\, M,  \nonumber\\
M' &\rightarrow& e^{-4i\nu}\, M'.
\end{eqnarray}
Therefore,  U(1)$_{\rm A}$ can only discriminate between two- and four-quark chiral nonets, but not among the three tetraquark substructures in (\ref{tetraq_abc}).  It is very desirable to explore whether there is a symmetry which might probe the individual  substructures of  (\ref{tetraq_abc}).  If so,  this symmetry should naturally be at the quark level and be cognizant to the SU(3) color dynamics.    Here we show that the symmetry presented in this work is in fact capable of discriminating the  three tetraquark substructures in (\ref{tetraq_abc}) in a manner that is consistent with the large $N_c$ approximation to QCD \cite{Witten}.

First, we note that the quark-antiquark scalar and pseudoscalar meson nonets are invariant under (\ref{res638274651}) and (\ref{res63827465}), respectively.   Let us denote the variation under (\ref{res638274651}) with subscript $1$, and that under (\ref{res63827465}) with subscript $2$. Then,
\begin{eqnarray}
&&\delta_1  S_a^b =
\delta_1[(q_{bA})^{\dagger}\gamma_4\frac{1}{2}q_{aA}]=k\partial_{\mu}[(q_{bA})^{\dagger}\gamma^4\gamma^{\mu}\frac{1}{2}q_{aA}]=0
\nonumber\\
&&i \delta_2 \phi_a^b =
\delta_2[(q_{bA})^{\dagger}\gamma_4\frac{\gamma_5}{2}q_{aA}]=ik\partial_{\mu}[(q_{bA})^{\dagger}\gamma^4\gamma^{\mu}\frac{\gamma_5}{2}q_{aA}]=0
\label{res53664}
\end{eqnarray}
Here we applied the conservation of the global currents.  These imply that the quark-antiquark chiral nonet is invariant under the following transformation:
\begin{eqnarray}
\delta M = \delta_1 S + i\delta_2 \phi = 0
\label{M_trans}
\end{eqnarray}
Note that  if we apply the transformation in Eq. (\ref{res638274651}) to the scalars (or fermion bilinears that do not contain $\gamma_5$) and that in Eq. (\ref{res63827465}) to pseudoscalars (or fermion bilinears that contain $\gamma_5$)  no significant cancellation occurs:
\begin{eqnarray}
&&\delta_2[(q_{bA})^{\dagger}\gamma_4\frac{1}{2}q_{aA}]=
\nonumber\\
&&k[i(\partial_{\mu}(q_{bA})^{\dagger})\gamma_4\gamma^{\mu}\frac{\gamma_5}{2}q_{aA}-(q_{bA})^{\dagger}\gamma_4\gamma^{\mu}\frac{\gamma_5}{2}\partial_{\mu}q_{aA}-
2g(q_{bA})^{\dagger}\gamma_4\gamma^{\mu}\frac{\gamma_5}{2}(t^m)_{AB}q_{aB}A^{m}_{\mu}]]
\nonumber\\
&&\delta_1[(q_{bA})^{\dagger}\gamma_4\frac{\gamma_5}{2}q_{aA}]=
\nonumber\\
&&k[(\partial_{\mu}(q_{bA})^{\dagger})\gamma_4\gamma^{\mu}\gamma_5\frac{1}{2}q_{aA}-(q_{bA})^{\dagger})\gamma_4\gamma^{\mu}\gamma_5\frac{1}{2}(\partial_{\mu}q_{aA})+
2ig(q_{bA})^{\dagger}\gamma_4\gamma^{\mu}\gamma_5\frac{1}{2}(t^m)_{AB}q_{aB}A^{m}_{\mu}]
\label{delta1phi_delta2S}
\end{eqnarray}

Based on (\ref{M_trans}),  it is evident that substructure ($a$) in (\ref{tetraq_abc}) is also invariant:
\begin{eqnarray}
\delta M_a^{(2)b}=\epsilon_{acd}\epsilon^{bef}(\delta (M^{\dagger})^c_e)(M^{\dagger})^d_f+\epsilon_{acd}\epsilon^{bef}(M^{\dagger})^c_e(\delta (M^{\dagger})^d_f)=0,
\label{res64557788}
\end{eqnarray}

Before we study the transformation of substructures ($b$) and ($c$) of (\ref{tetraq_abc})  under the fermion symmetries
(\ref{res638274651}) and (\ref{res63827465}), we first note that by applying Fierz transformations we can rewrite each of these substructures as a linear combination of a ``molecular'' piece and a diquark-antidiquark piece \cite{Jora3}:

	\begin{eqnarray}
	(b) \hskip .5cm M_g^{(3)f}&=&(L^{gA})^{\dagger}R^{fA}
	\nonumber\\
	&=&2\epsilon_{gab}\epsilon^{fde}\Bigg[\bar{q}_{aA}\frac{1-\gamma_5}{2}q_{dA}\bar{q}_{bB}\frac{1-\gamma_5}{2}q_{eB}-\bar{q}_{aA}\frac{1-\gamma_5}{2}q_{dB}\bar{q}_{bB}\frac{1-\gamma_5}{2}q_{eA}\Bigg]
	\nonumber\\
	&=& 2 M^{(2) f}_g - 2 {\widehat M}^f_g
	\nonumber\\
	\nonumber\\
    (c) \hskip .5cm M_g^{(4)f}&=&(L^{g}_{\mu\nu,AB})^{\dagger}R^{f}_{\mu\nu,AB}
	\nonumber\\
	&=&-4\epsilon_{gab}\epsilon^{fde}\Bigg[\bar{q}_{aA}\frac{1-\gamma_5}{2}q_{dA}\bar{q}_{bB}\frac{1-\gamma_5}{2}q_{eB}+\bar{q}_{aA}\frac{1-\gamma_5}{2}q_{dB}\bar{q}_{bB}\frac{1-\gamma_5}{2}q_{eA}\Bigg]
	\nonumber\\
	&=& -4  M^{(2) f}_g - 4 {\widehat M}^f_g .
	\label{teraq43524}
	\end{eqnarray}
Using (\ref{res64557788}),  we find that
\begin{eqnarray}
\delta M_g^{(3)f} &=& -2 \delta {\widehat M}_g^{(3)f} \nonumber \\
\delta M_g^{(4)f} &=& - 4 \delta {\widehat M}_g^{(3)f}
\end{eqnarray}
where
\begin{eqnarray}
\delta {\widehat M}_g^{(3)f} &=&  \delta(\bar{q}_{aA}\frac{1-\gamma_5}{2}q_{dB}\bar{q}_{bB}\frac{1-\gamma_5}{2}q_{eA})
\nonumber\\
&=&\Bigg[\delta_1(\bar{q}_{aA}\frac{1}{2}q_{dB})+\delta_2(\bar{q}_{aA}\frac{-\gamma_5}{2}q_{dB})\Bigg](\bar{q}_{bB}\frac{1-\gamma_5}{2}q_{eA})
\nonumber \\
&&+
(\bar{q}_{aA}\frac{1-\gamma_5}{2}q_{dB})\Bigg[\delta_1(\bar{q}_{bB}\frac{1}{2}q_{eA})+\delta_2(\bar{q}_{bB}\frac{-\gamma_5}{2}q_{eA})\Bigg]
\label{transform75664}
\end{eqnarray}
However the variations,
\begin{eqnarray}
&&\delta_1(\bar{q}_{aA}\frac{1}{2}q_{dB})=\frac{1}{2}\Bigg[\partial_{\rho}(\bar{q}_{aA}\gamma^{\rho}q_{dB})+
ig\bar{q}_{aC}(t^m)_{CA}\gamma^{\rho}q_{dB}A^m_{\rho}-ig\bar{q}_{aA}(t^m)_{BC}\gamma^{\rho}q_{dC}A^m_{\rho}\Bigg]
\nonumber\\
&&\delta_2(\bar{q}_{aA}\frac{\gamma_5}{2}q_{dB})=\frac{1}{2}\Bigg[\partial_{\rho}(\bar{q}_{aA}\gamma^{\rho}q_{dB})-
g\bar{q}_{aC}(t^m)_{CA}\gamma^{\rho}q_{dB}A^m_{\rho}+g\bar{q}_{aA}(t^m)_{BC}\gamma^{\rho}q_{dC}A^m_{\rho}\Bigg].
\label{resq218455}
\end{eqnarray}
cannot cancel because different color indices appear in the respective terms.    Therefore, in summary:
\begin{eqnarray}
\delta M_g^{(2)f} &=& 0 \nonumber \\
\delta M_g^{(3)f} &\ne & 0\nonumber \\
\delta M_g^{(4)f} &\ne& 0
\label{delta_M234}
\end{eqnarray}
This means that the ``molecular'' structure ($a$) in (\ref{teraq43524}) is the favored structure with respect to the new fermion  symmetry discussed in this work.   This is consistent with large $N_c$ approximation of QCD, because ${\widehat M}$ is of order $1/N_c$ (see \cite{Witten,Manohar}) and therefore:
\begin{eqnarray}
M_g^{(3)f} &=& 2 M_g^{(2)f} + {\cal O} {1\over N_c} \nonumber \\
M_g^{(4)f} &=& -4 M_g^{(2)f} + {\cal O} {1\over N_c}
\end{eqnarray}
which means in $N_c\rightarrow \infty$ limit tetraquarks approach  ``molecular'' structure in agreement with the present symmetry which favors ``molecular'' structure (\ref{delta_M234}).

The fact that this symmetry applies differently to quark constituents in different composite fields can be understood based on the linear sigma model with a partition function where the fields of integration are quark states instead of meson ones. Then one can separate the quarks that belong to the scalar structures ($q_i$) from the quarks ( $q_i'$) that belong to the pseudoscalar one (or any generalization of this) as follows:
\begin{eqnarray}
Z=\int d q_i d \bar{q}_i d q'_i d\bar{q}'_i \delta(q_i-q_i')\delta(\bar{q}_i-\bar{q}_i')\exp[\int d^4 x {\cal L}].
\label{lagr65774756}
\end{eqnarray}
We apply the transformation in Eq. (\ref{res638274651}) to $q_i$ and $\bar{q}_i$ and that in Eq. (\ref{res63827465}) to the quarks $q_i'$ and $\bar{q}_i'$. Furthermore we apply the Fujikawa method and the delta function to determine that the corresponding Jacobian leads to an anomaly. The fact that the symmetry is anomalous is not new as from the start we knew that the axial symmetry was anomalous. The net result is that the quark Lagrangian is invariant up to this anomaly term where the two species (prime and unprimed) of quarks converge into a single one as they should by applying the delta function.

 According to our discussion at the end of section II we expect that the axial anomaly will be present in the low-energy QCD effective Lagrangian. The axial $U(1)_{\rm A}$  anomaly term for the color group has been treated in detail within a generalized linear sigma model in \cite{Jora1}-\cite{Jora4}  and most recently in \cite{Jora5} for the electromagnetic one.  For example,  the color $U(1)_{\rm A}$ anomaly introduces in the linear sigma model of   low-energy QCD an effective term of the type:
\begin{eqnarray}
{\cal L}_a\propto\epsilon^{\mu\nu\rho\sigma}F^a_{\mu\nu}F^a_{\rho\sigma}[\ln(\det M)-\ln(\det M^{\dagger})].
\label{res534243}
\end{eqnarray}
Here $M$ is the chiral noent defined in (\ref{non546788}). Upon scalars developing  a VEV  ($S_0$)  and spontaneously breaking chiral symmetry, the determinant in Eq. (\ref{res534243}) can be expanded to get (in second order) also a term of the type:
\begin{eqnarray}
{\cal L}_a\propto...+\epsilon^{\mu\nu\rho\sigma}F^a_{\mu\nu}F^a_{\rho\sigma}\frac{1}{2S_0^2}i{\rm Tr}[\Phi]{\rm Tr}[S]+...
\label{expuutr}
\end{eqnarray}
Then apart from the particular flavor structure which is not essential for our discussion we get exactly the term that we expect to contribute to the anomaly in Eq. (\ref{finalres73546789}).

To conclude,  the symmetries introduced and discussed here should be by themselves symmetries of a complete effective Lagrangian that contains all possible states including vector and tensor ones and may contain hints not only on how to build an appropriate model  but also for the actual hadron structure and composition.

\section*{Acknowledgments} \vskip -.5cm

A. H. F gratefully acknowledges the support of this research by the College of Arts and Sciences of SUNY Poly in the Spring semester of 2017.

\end{document}